\documentclass[12pt,preprint]{aastex}
\usepackage{apjfonts, emulateapj5, natbib}
\input psfig.tex

\def\aa{{A\&A}}

\def\aj{{AJ}}

\def\annrev{{ARA\&A}}
\def\apj{{ApJ}}
\def\apjs{{ApJS}}
\def\asec{$^{\prime\prime}$}

\def\ha{H$\alpha$}

\def\hb{H$\beta$}

\def\kms{km s$^{-1}$}
\def\lamb{$\lambda$}
\def\lax{{$\mathrel{\hbox{\rlap{\hbox{\lower4pt\hbox{$\sim$}}}\hbox{$<$}}}$}}
\def\gax{{$\mathrel{\hbox{\rlap{\hbox{\lower4pt\hbox{$\sim$}}}\hbox{$>$}}}$}}
\def\simlt{\lower.5ex\hbox{$\; \buildrel < \over \sim \;$}}
\def\simgt{\lower.5ex\hbox{$\; \buildrel > \over \sim \;$}}
\def\lum{erg s$^{-1}$}
\def\mbh{{$M_{\rm BH}$}}
\def\micron{{$\mu$m}}
\def\mnras{{MNRAS}}

\def\percm2{cm$^{-2}$}
\def\peryr{yr$^{-1}$}

\def\solmass{$M_\odot$}
\def\oii{[\ion{O}{2}]}

\def\caii{\ion{Ca}{2}}
\def\feii{\ion{Fe}{2}}

\def\hii{\ion{H}{2}}
\def\oiii{[\ion{O}{3}]}

\def\nii{[\ion{N}{2}]}

\def\sii{[\ion{S}{2}]}

\def\mbh{{$M_{\rm BH}$}}
\def\msig{{$M_{\rm BH}-\sigma_*$}}

\newdimen\digitwidth      
\setbox1=\hbox{0}       
\digitwidth=\wd1        
\catcode`"=\active      
\def"{\kern\digitwidth}

\slugcomment{To appear in {\it The Astrophysical Journal.}}
\lefthead{Ho \& Kim}
\righthead{}

\begin{document}

\title{Low-mass Active Galactic Nuclei with Rapid X-ray Variability}

\author{Luis C. Ho\altaffilmark{1,2} and Minjin Kim\altaffilmark{3,4}}

\altaffiltext{1}{Kavli Institute for Astronomy and Astrophysics, Peking
University, Beijing 100871, China}

\altaffiltext{2}{Department of Astronomy, School of Physics, Peking University, Beijing 100871, China}

\altaffiltext{3}{Korea Astronomy and Space Science Institute, Daejeon 305-348,
Republic of Korea}

\altaffiltext{4}{University of Science and Technology, Daejeon 305-350,
Republic of Korea}

\begin{abstract}
We present a detailed study of the optical spectroscopic properties of 12 
active galactic nuclei (AGNs) with candidate low-mass black holes (BHs) 
selected by Kamizasa et al. through rapid X-ray variability. The high-quality, 
echellette Magellan spectra reveal broad \ha\ emission in all the sources, 
allowing us to estimate robust viral BH masses and Eddington ratios for this
unique sample.  We confirm that the sample contains low-mass BHs accreting at 
high rates: the median $M_{\rm BH} = 1.2\times 10^6$ \solmass\ and median 
$L_{\rm bol}/L_{\rm Edd}=0.44$.   The sample follows the $M_{\rm BH}-\sigma_*$ 
relation, within the considerable scatter typical of pseudobulges, the probable
hosts of these low-mass AGNs.  Various lines of evidence suggest that ongoing 
star formation is prevalent in these systems.  We propose a new strategy to 
estimate star formation rates in AGNs hosted by low-mass, low-metallicity 
galaxies, based on modification of an existing method using the strength of 
\oii\ \lamb 3727, \oiii\ \lamb 5007, and X-rays.
\end{abstract}


\section{Introduction}

Central black holes (BHs) constitute an integral component of the nuclear 
regions of galaxies.  Much progress has been made in the detection and 
characterization of the statistical properties of supermassive BHs in nearby 
galaxies, over the mass range \mbh\ $\approx 10^6 - 10^{10}$ \solmass\ (e.g., 
G\"ultekin et al. 2009; McConnell \& Ma 2013; Kormendy \& Ho 2013).  During 
the last 10 years, there has also been a number of successful attempts to 
identify the rarer population of lower mass central BHs, with \mbh\ \lax\ 
$10^6$ \solmass, in bands ranging from the optical (e.g., Greene \& Ho 2004, 
2007c; Barth et al. 2008; Seth et al. 2010; Dong et al. 2012b; Reines et al. 
2013; Moran et al. 2014; Baldassare et al. 2015), to the radio (Reines et al. 
2011, 2014), and the mid-infrared (Satyapal et al. 2008, 2009, 2014; Goulding 
\& Alexander 2009).  X-ray images, particularly at high angular resolution, 
can effectively identify candidate active galactic nuclei (AGNs) as compact 
nuclear cores in low-mass, late-type---even dwarf---galaxies (Desroches \& Ho 
2009; Zhang et al. 2009; Reines et al. 2011, 2014; Araya~Salvo et al. 2012; 
Schramm et al. 2013; Lemons et al. 2015).  

Exploiting the fact that the amplitude of X-ray variability in AGNs 
scales inversely with BH mass (e.g., Lu \& Yu 2001; Papadakis 2004; McHardy 
et al. 2006), Kamizasa et al. (2012) used the Second {\it XMM-Newton}\ 
Serendipitous Source Catalogue (Watson et al. 2009) to select a sample of 16 
AGNs with candidate low-mass BHs, based on the presence of rapid variability in 
their X-ray light curve.  For the 15 sources for which they could compute 
the  ``normalized excess variance'' in the 0.5--10 keV band, they estimate 
that their sample contains \mbh\ $\approx (1.1-6.6) \times 10^6$ \solmass.  The
subset with available redshift estimates has an inferred median Eddington 
ratio of $\lambda_{\rm E} \equiv L_{\rm bol}/L_{\rm Edd} = 0.24$, where 
$L_{\rm bol}$ is the bolometric luminosity and $L_{\rm Edd}$ is the Eddington 
luminosity.  The remaining object, 2XMM~J123103.2+110648, has an exceptionally 
prominent soft X-ray excess, which Terashima et al. (2012) model with a 
multicolor disk blackbody with an inner disk temperature of $kT = 0.16-0.21$ 
keV.  In a follow-up optical study, Ho et al. (2012) confirm the AGN nature of 
the source but find that it emits only narrow emission lines, an unexpected 
result given the lack of significant intrinsic X-ray absorption and the high 
Eddington ratio ($\lambda_{\rm E}$ \gax\ 0.5) derived from its estimated BH 
mass of $\sim 10^5$ \solmass.

The majority of Kamizasa et al.'s sample has no previous optical spectroscopic 
data, not even basic information such as spectroscopic redshifts, making it 
impossible to derive many fundamental parameters for this unique sample.  
Moreover, in light of the intriguing properties of 2XMM~J123103.2+110648, it 
would be of interest to investigate the optical spectral properties of these 
objects more systematically.  This paper reports detailed high-resolution
, high--signal-to-noise ratio (S/N) optical spectra for 12 of the 15 sources 
from the sample of Kamizasa et al. (2012) with BH masses previously estimated 
through X-ray variability.  As the sample was initially designed to select 
low-mass objects, we anticipate the emission lines, both broad and narrow, to 
have small velocity widths; moderately high dispersion would be required to 
measure useful kinematics.

\section{Observations, Data Reduction, and Spectral Fits}

The observations, summarized in Table~1, were obtained with the Magellan 6.5m 
Clay telescope at Las Campanas Observatory, during the course of two observing 
runs in 2013, using the Magellan Echellette (MagE) spectrograph (Marshall et 
al. 2008).  The spectra cover $\sim$3100 \AA\ to 1 \micron\ in 15 spectral 
orders at a resolving power for a 1\asec\ slit, as determined from the sky 
emission lines, of $\lambda/\Delta \lambda = 4627$, which corresponds to an 
instrumental velocity dispersion of $\sigma_i = 27.6$ \kms.  The brightnesses
of the sources range from $g \approx 18$ to 21 mag.  We observed featureless 
standard stars for flux calibration and telluric correction, as well as 
late-type stellar templates for velocity dispersion measurement. The observing 
conditions were mostly photometric, but the observations were occasionally 
carried out under thin cirrus.  The seeing varied from 0\farcs6 to 1\farcs2. 

\begin{figure*}[t]
\centerline{\psfig{file=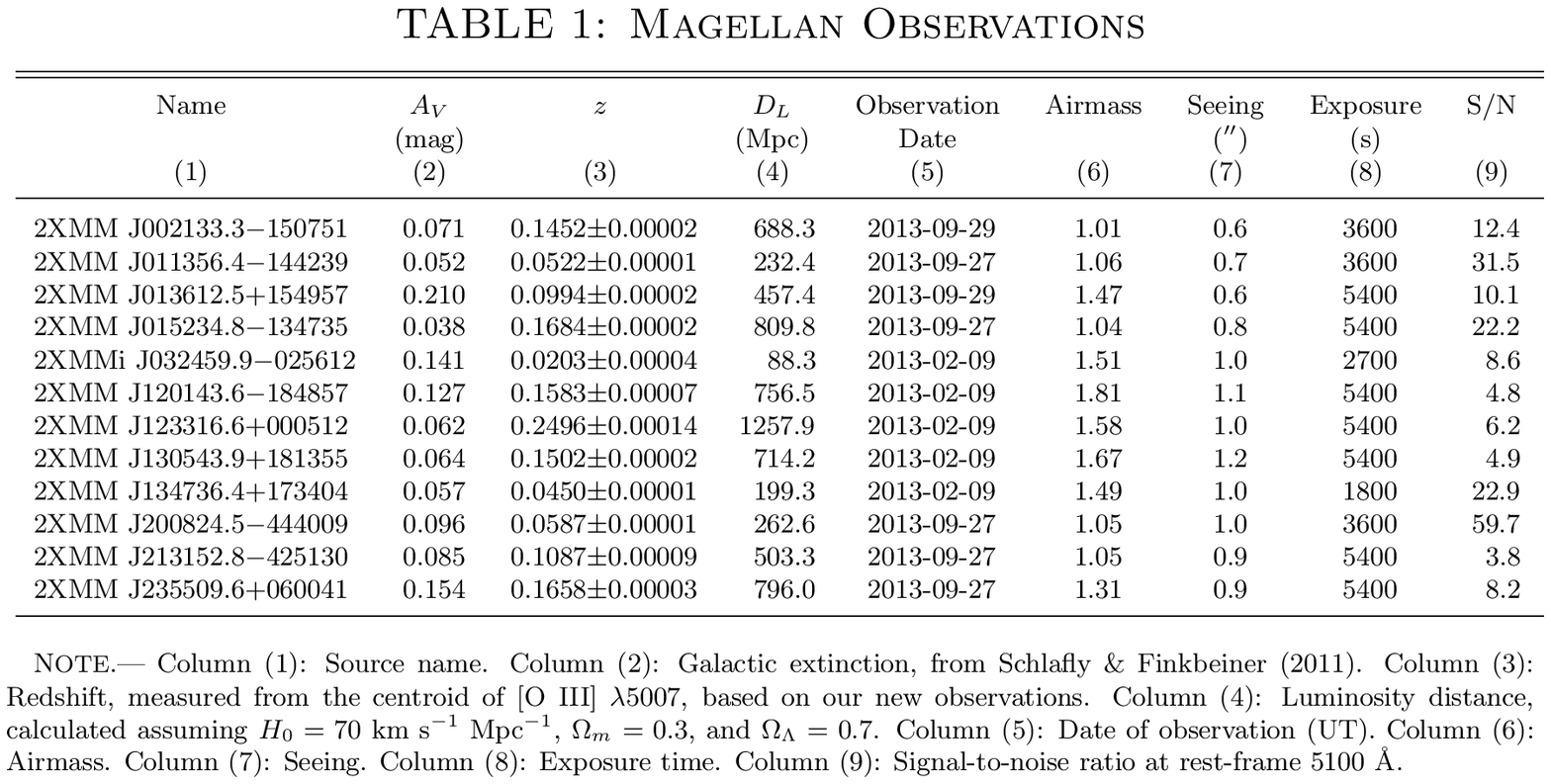,width=18.5cm,angle=0}}
\end{figure*}

\begin{figure*}[t]
\centerline{\psfig{file=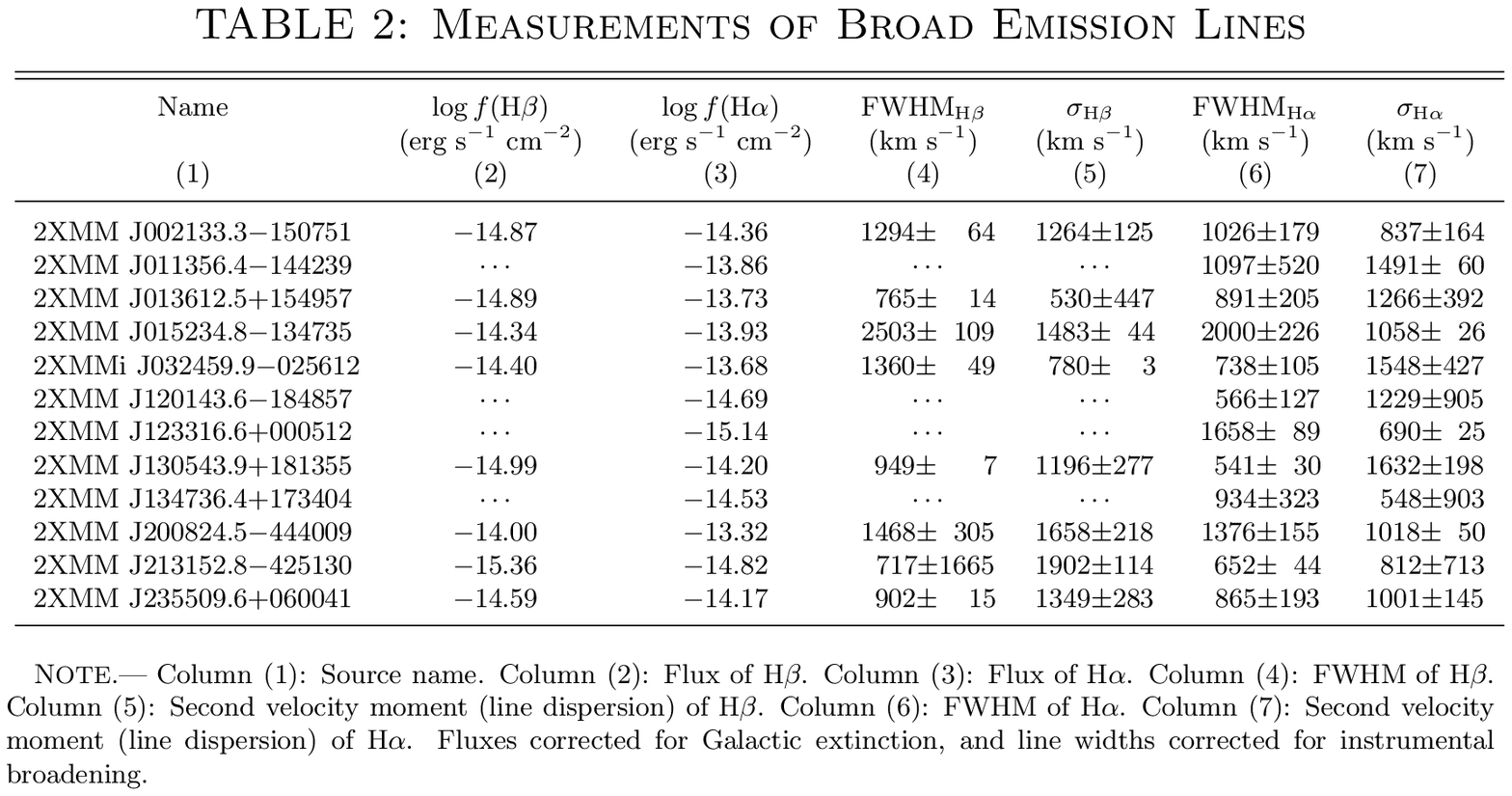,width=18.5cm,angle=0}}
\end{figure*}

\noindent
The slit 
was aligned along the parallactic angle to minimize slit losses due 
to differential atmospheric refraction.  

The data were reduced using the IDL routine {\tt mage\_reduce} developed and 
kindly provided by George Becker. We performed flat-fielding correction using 
twilight flats and contemporaneous dome flats taken just after each science
exposure.  Wavelength calibration was done using ThAr arc spectra.  After 
extracting 1-D spectra, we apply flux calibration and telluric correction.  We 
correct for Galactic extinction using the extinction values of Schlafly \& 
Finkbeiner (2011) and the reddening curve of Cardelli et al. (1989).  

Figure~1 shows the final spectra for the sample.  Strong emission lines, and 
in some cases a featureless continuum, are present in all the sources, 
superposed on a weak stellar continuum characteristic of intermediate-age 
stars.  The strong emission lines, particularly \oiii\ \lamb 5007, allow us to
determine--- accurately and for the first time in most cases---the redshifts 
of the sample (Table~1).  We decompose the spectra into host galaxy starlight, 
a featureless AGN power-law continuum, \feii\ pseudocontinuum, and a residual, 
pure emission-line spectrum, closely following the treatment described in Ho 
\& Kim (2009).  We perform a multi-component Gaussian fit of the 
continuum-subtracted spectrum to derive fluxes, widths, and radial velocities 
of the major broad and narrow emission lines (see bottom panels of each object 
in Figure~1).  To isolate the broad \ha\ line, we follow previous practice 
(e.g., Ho et al. 1997) and constrain the profile of \nii\ \lamb\lamb 6548,~6583
and the narrow component of \ha\ to that of \sii\ \lamb\lamb 6716,~6731.  
Likewise, we tie the narrow component of \hb\ to \oiii\ \lamb 5007.  Each of 
the lines of the \sii\ doublet usually can be fit adequately with a single 
Gaussian, whereas \oiii\ requires an extended, often blueshifted, wing 
component in addition to a Gaussian core.  The final, 

\clearpage
\vskip 0.3cm
\begin{figure*}[t]
\figurenum{1}
\centerline{\psfig{file=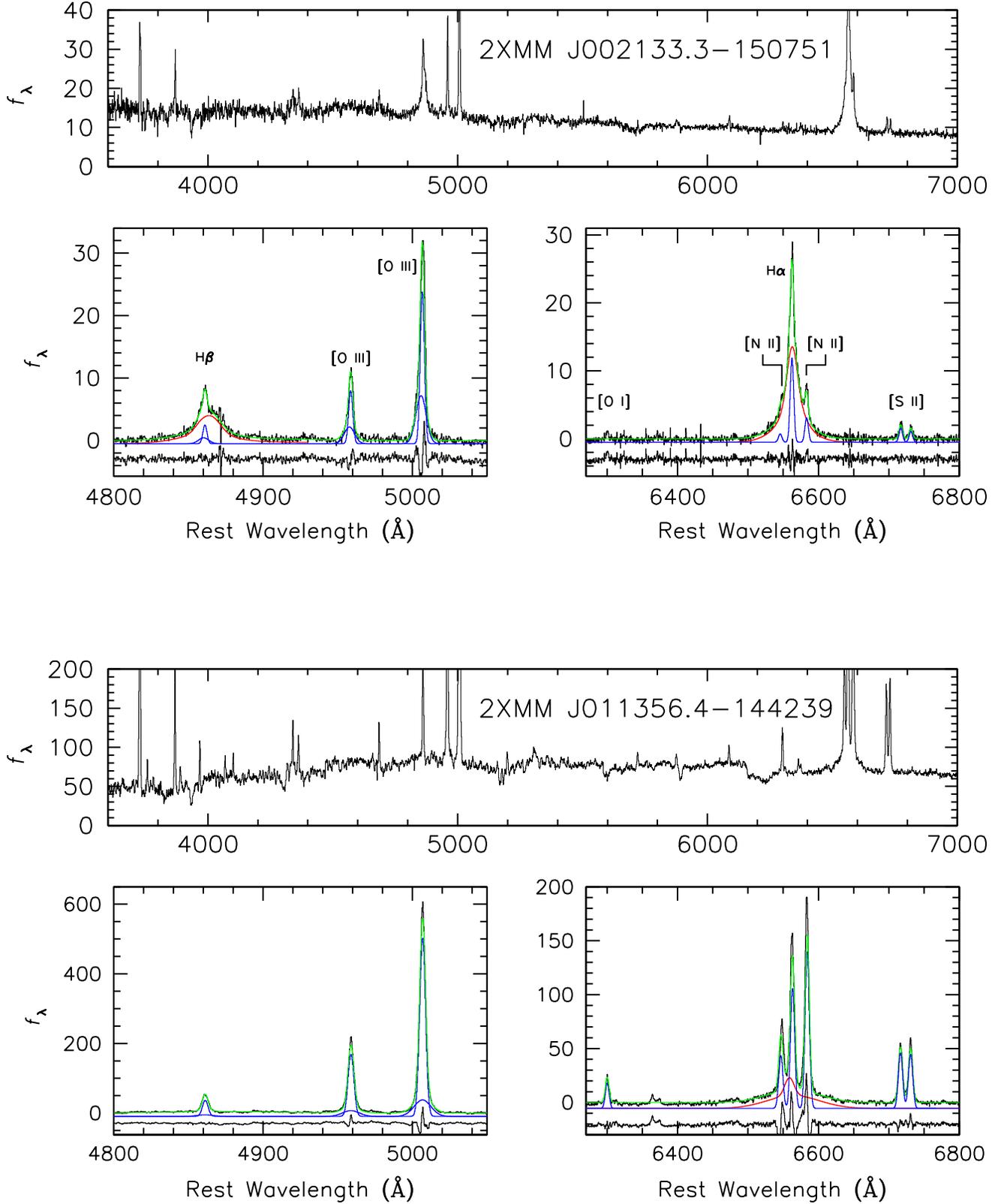,width=17.5cm,angle=0}}
\figcaption[fig1.ps]{
Magellan/MagE spectra of the sample, shifted to the rest-frame and corrected
for Galactic extinction.  The top panel focuses on the broad-band continuum
features; the spectra have been slightly binned for the purposes of the display.
The bottom panels highlight the major emission lines of interest.
The black histogram is the observed emission-line spectrum, after
subtraction of a featureless AGN continuum, \feii\ emission, and host galaxy
starlight.  Multi-component decomposition of the emission lines are shown
in blue for the narrow lines and in red for broad \ha\ and \hb\ (offset 
 vertically by an arbitrary constant for clarity); the final 
model is plotted in green.  The bottom spectrum shows the residuals of the 
fit, also offset slightly vertically for clarity.
\label{fig1}}
\end{figure*}
\vskip 0.3cm

\clearpage
\vskip 0.3cm
\begin{figure*}[t]
\figurenum{1}
\centerline{\psfig{file=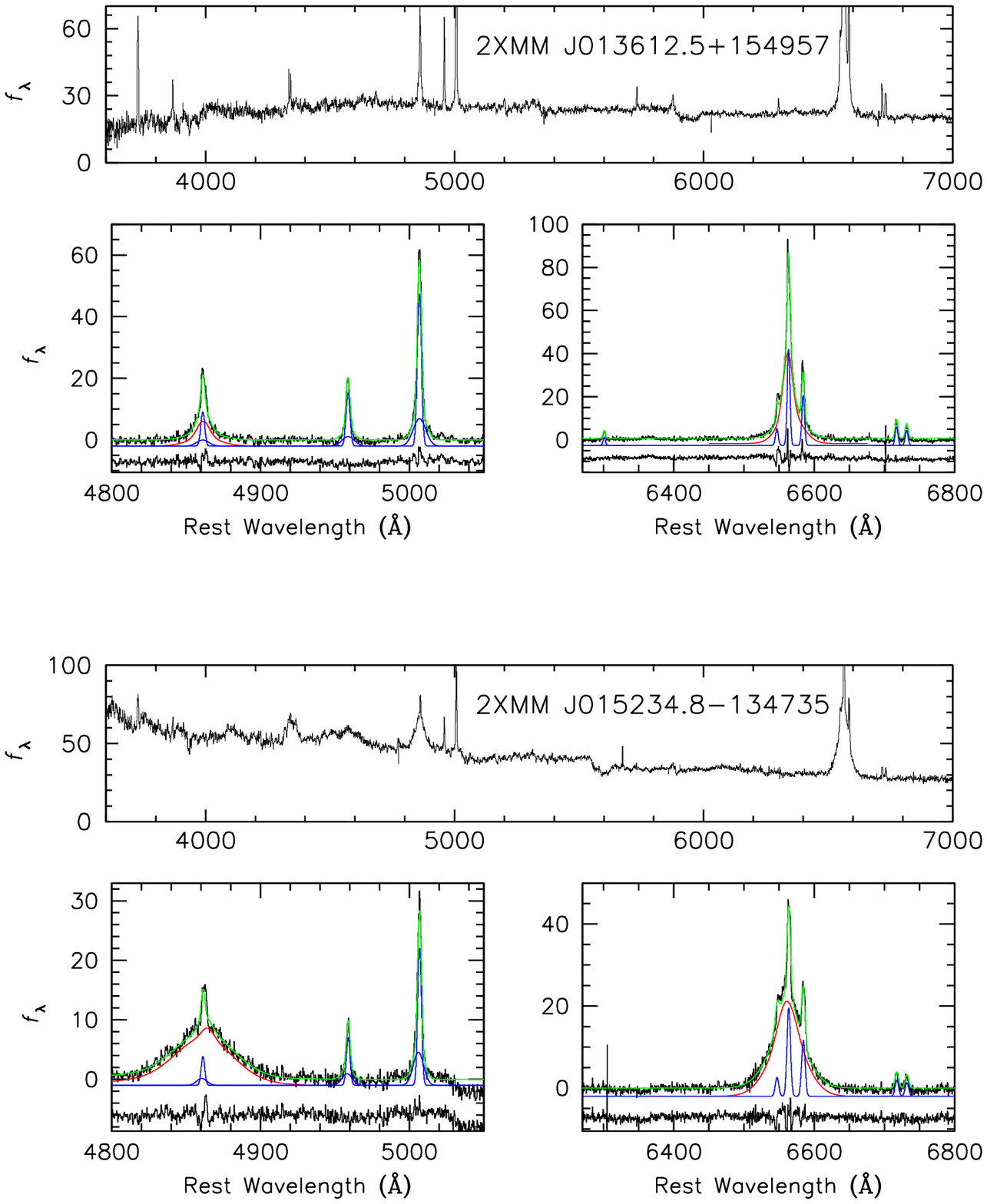,width=17.5cm,angle=0}}
\figcaption[fig1.ps]{
Figure 1, continued.
\label{fig1}}
\end{figure*}
\vskip 0.3cm

\clearpage
\vskip 0.3cm
\begin{figure*}[t]
\figurenum{1}
\centerline{\psfig{file=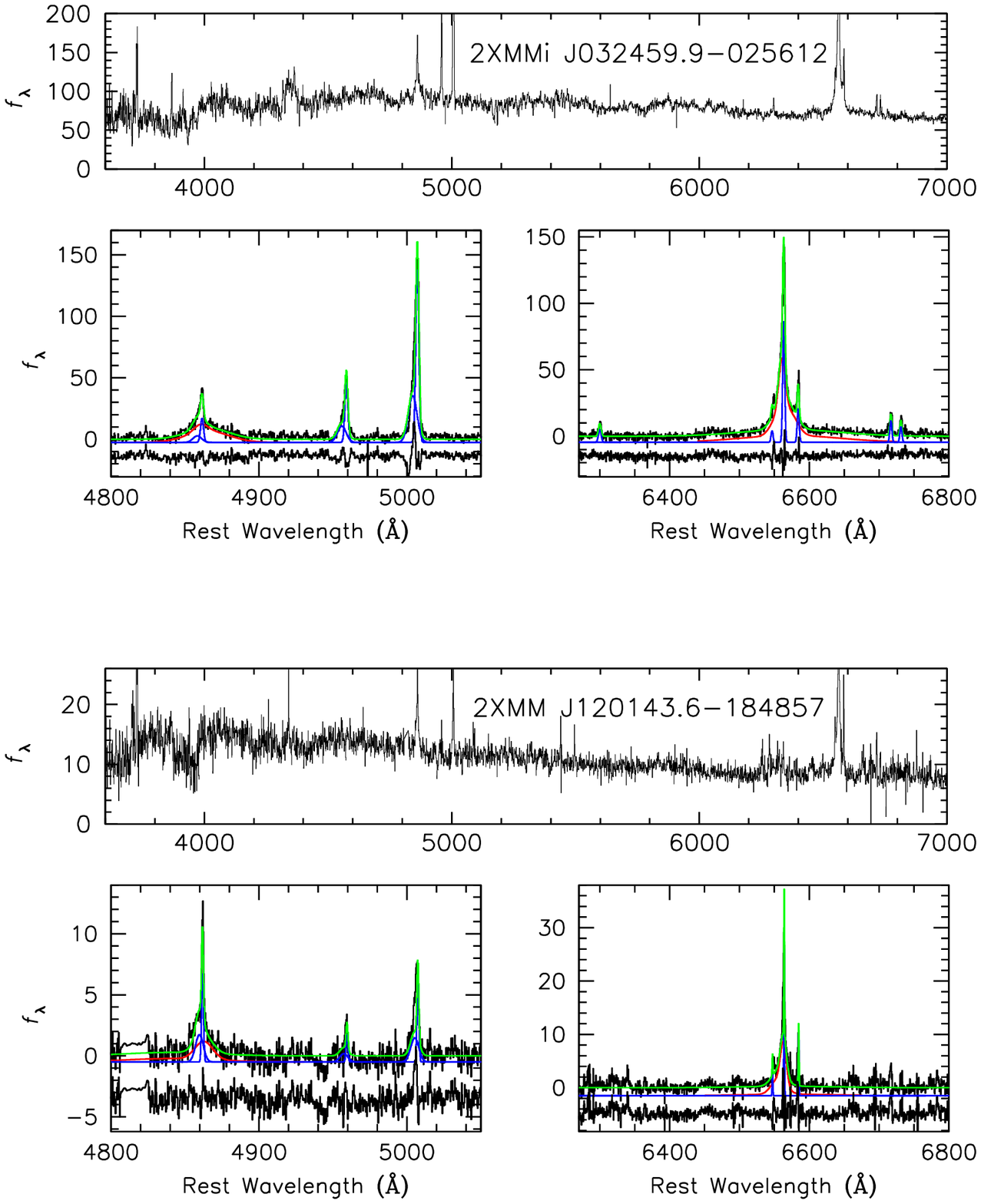,width=17.5cm,angle=0}}
\figcaption[fig1.ps]{
Figure 1, continued.
\label{fig1}}
\end{figure*}
\vskip 0.3cm

\clearpage
\vskip 0.3cm
\begin{figure*}[t]
\figurenum{1}
\centerline{\psfig{file=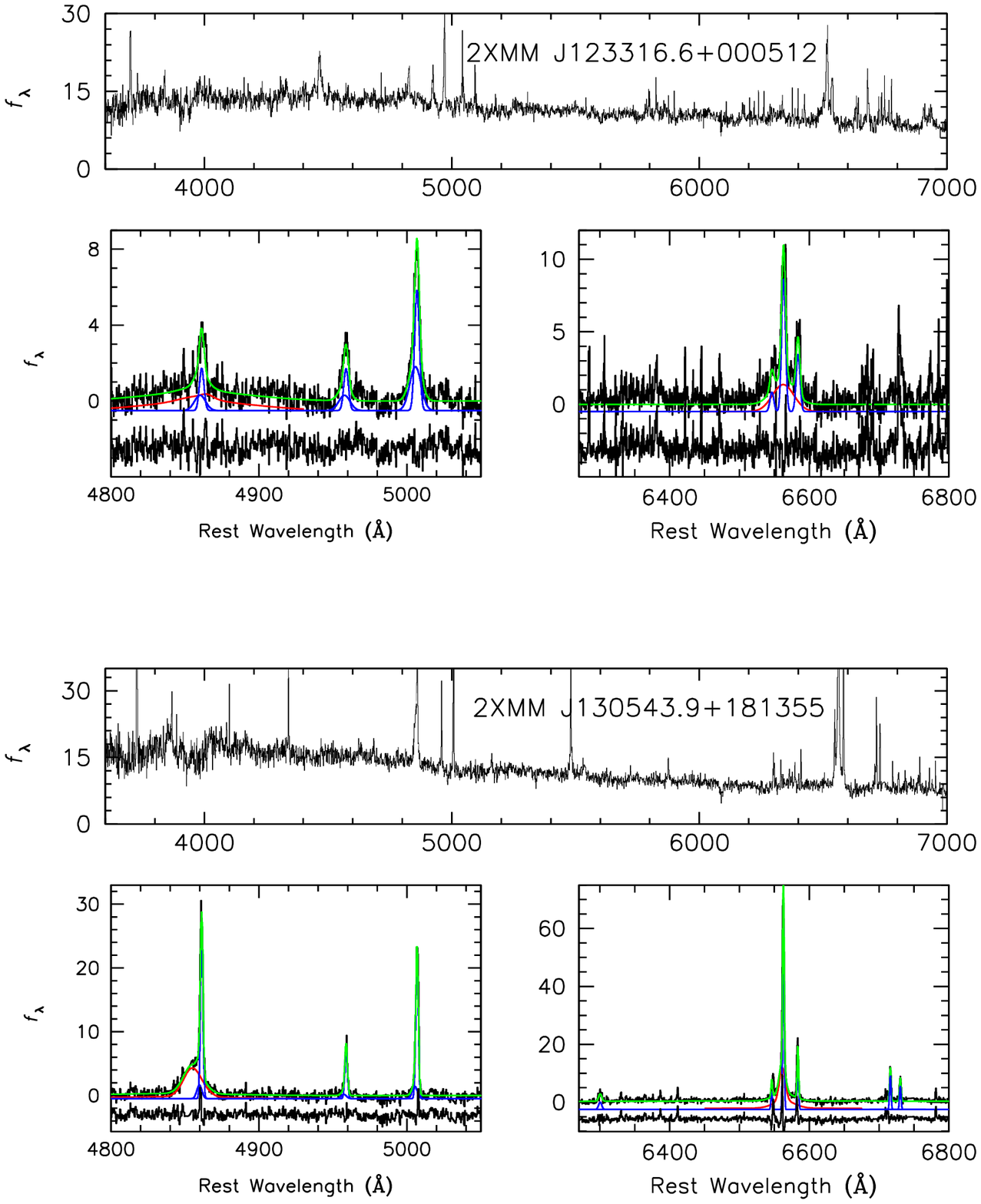,width=17.5cm,angle=0}}
\figcaption[fig1.ps]{
Figure 1, continued.
\label{fig1}}
\end{figure*}
\vskip 0.3cm

\clearpage
\vskip 0.3cm
\begin{figure*}[t]
\figurenum{1}
\centerline{\psfig{file=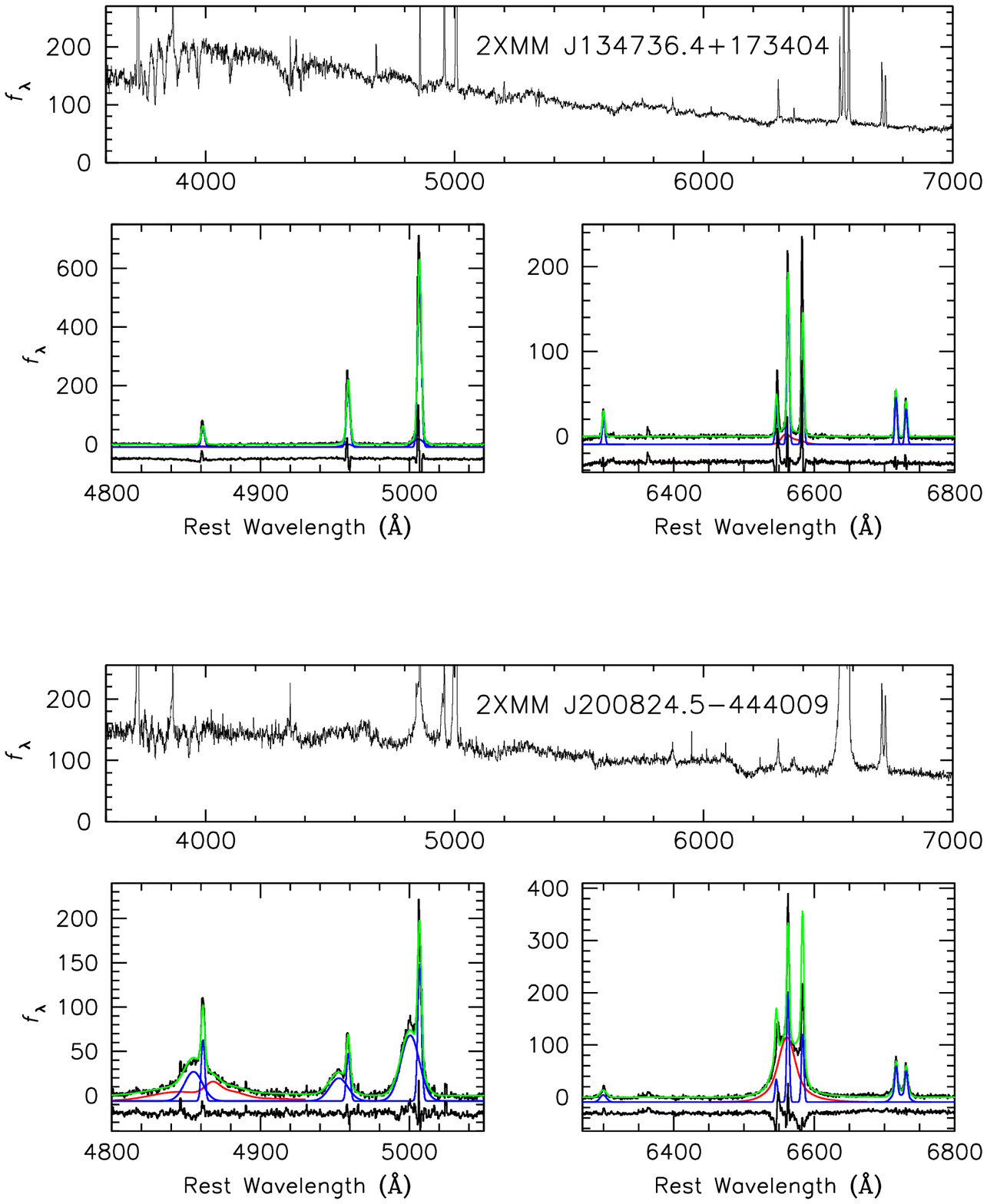,width=17.5cm,angle=0}}
\figcaption[fig1.ps]{
Figure 1, continued.
\label{fig1}}
\end{figure*}
\vskip 0.3cm

\clearpage
\vskip 0.3cm
\begin{figure*}[t]
\figurenum{1}
\centerline{\psfig{file=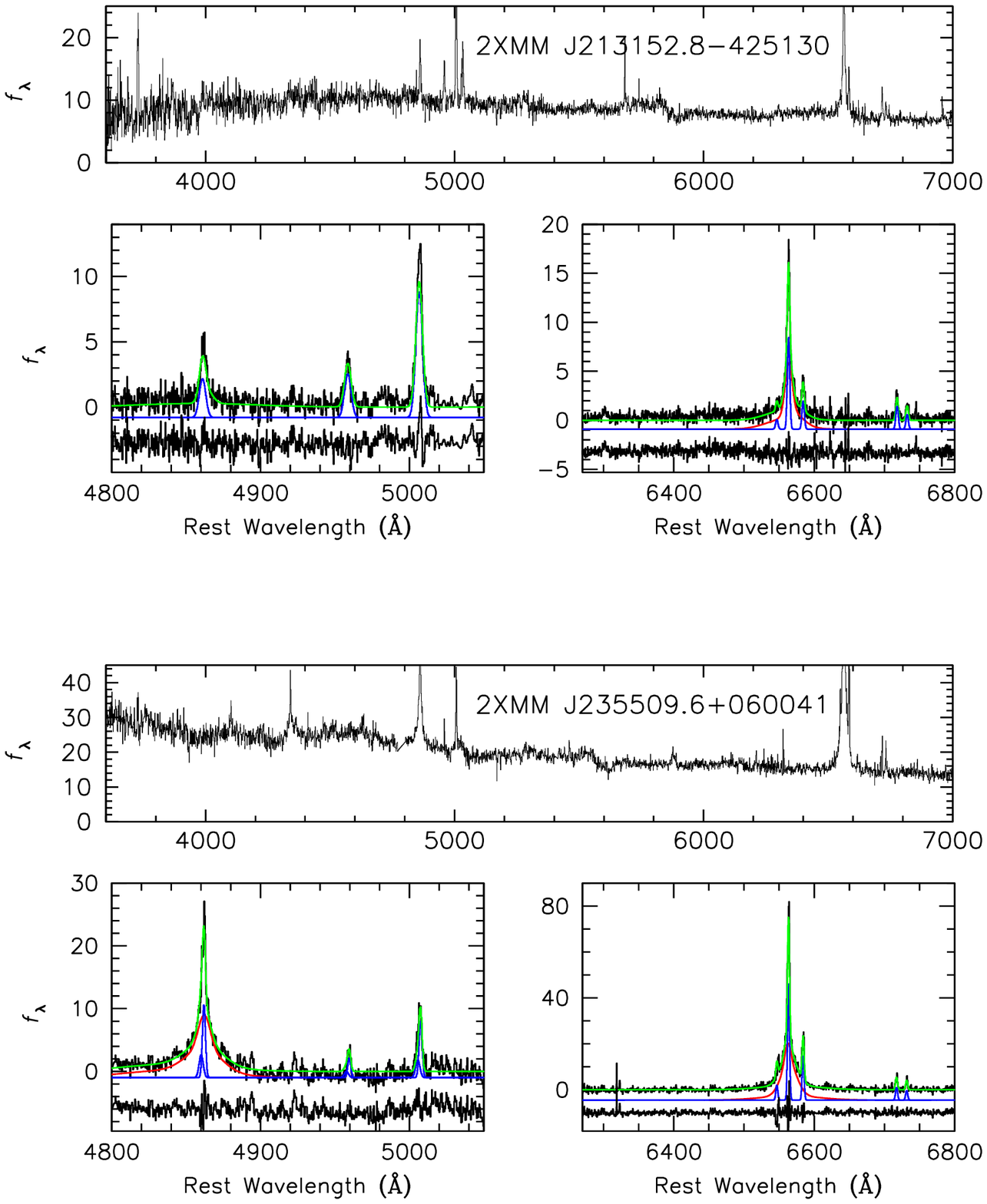,width=17.5cm,angle=0}}
\figcaption[fig1.ps]{
Figure 1, continued.
\label{fig1}}
\end{figure*}
\vskip 0.3cm
\clearpage

\vskip 0.3cm
\psfig{file=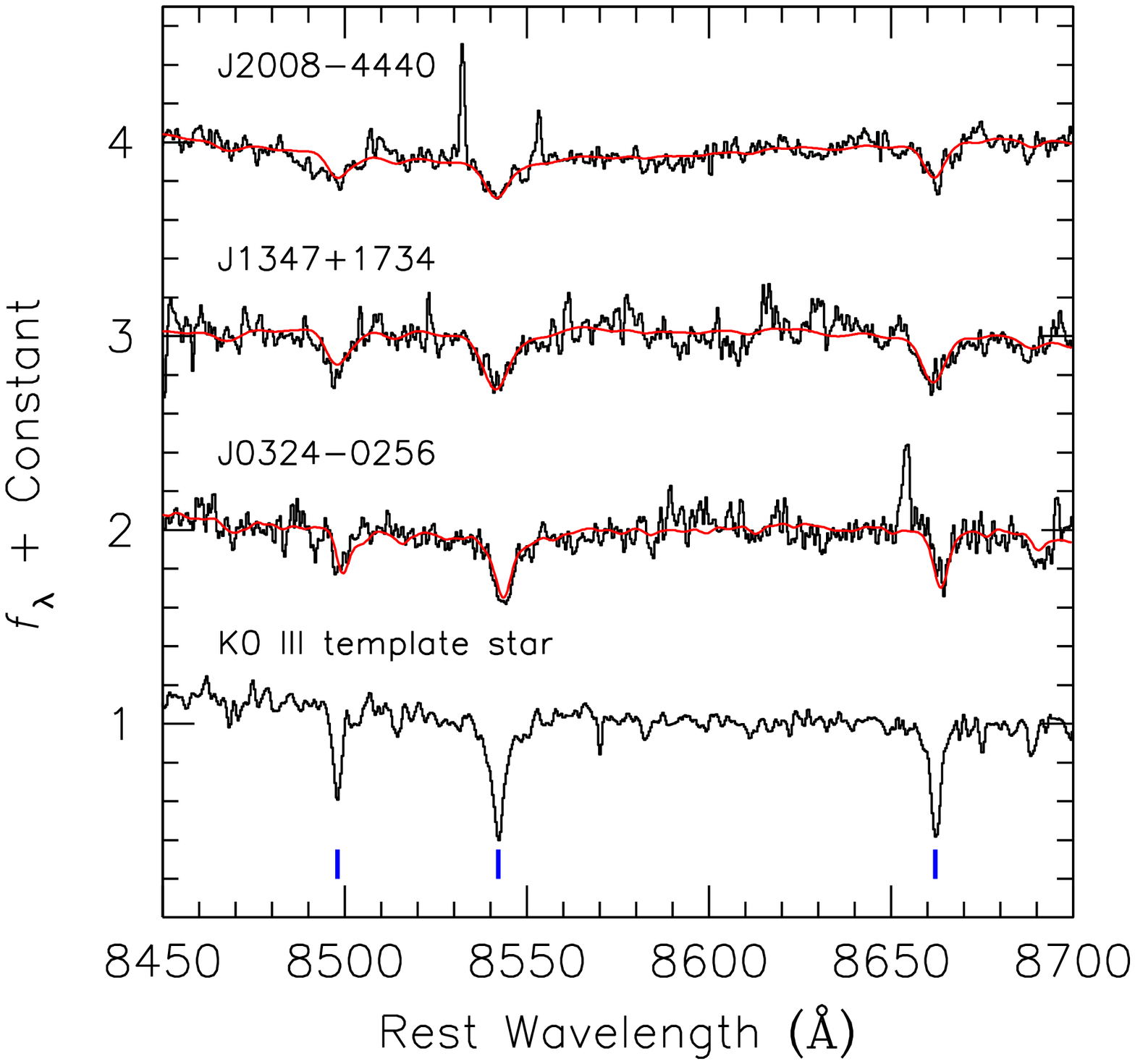,width=8.75cm,angle=0}
\figurenum{2}
\vskip 0.3cm
\figcaption[fig2.ps]{
Detection of the \caii\ IR triplets in 2XMMi~J032459.9$-$025612,
2XMM~J134736.4+173404, and 2XMM~J200824.5$-$444009.  The black histograms plot
the original data, and the red line is the best-fit spectrum of a broadened
K0~III template star, which is shown on the bottom.  The spectra have been
normalized to unity and shifted vertically for the purposes of the display.
Blue vertical tickmarks identify the \caii\ IR triplets.
\label{fig2}}
\vskip 0.3cm

\noindent
fitted parameters for 
the broad and narrow emission lines are listed in Tables~2 and 3, respectively.

The \caii\ infrared (IR) triplets (\lamb\lamb\lamb 8498, 8542, 8662) are ideal 
features for deriving stellar velocity dispersions, especially for composite 
stellar populations and AGN host galaxies, because of their relative 
insensitivity to template mismatch (Dressler 1984) and reduced contamination 
from nonstellar emission (Barth et al. 2002; Greene \& Ho 2006a).  The three 
sources with the lowest redshifts cover the \caii\ IR triplets in 

\psfig{file=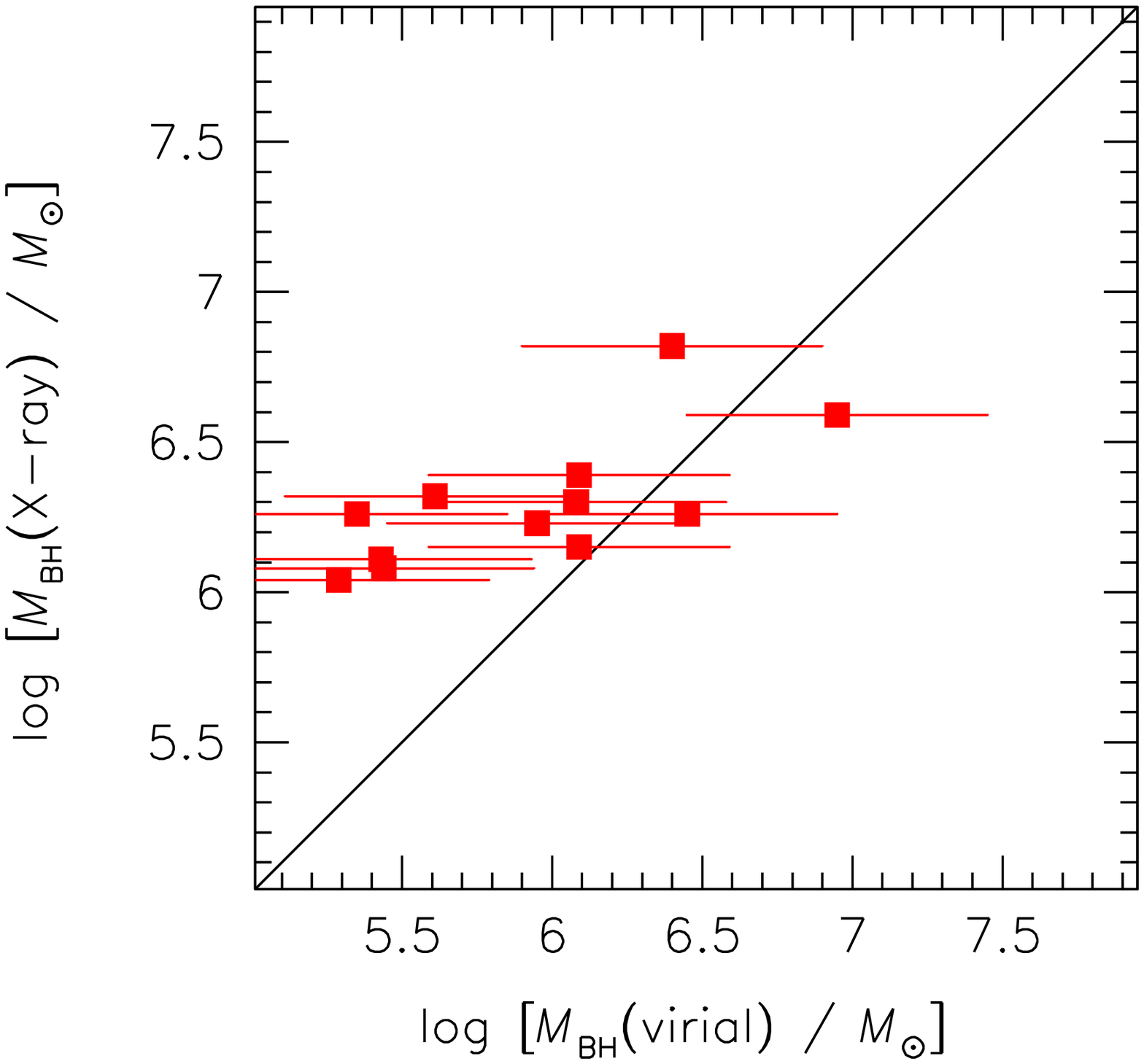,width=8.75cm,angle=0}
\figurenum{3}
\vskip 0.3cm
\figcaption[fig3.ps]{Comparison of the new virial BH masses derived from
our new detections of broad \ha\ emission with the BH masses estimated by
Kamizasa et al. (2012) based on X-ray variability.  The solid line denotes the
1:1 relation. The virial BH masses are assumed to have uncertainties of 0.5 dex.
\label{fig3}}

\psfig{file=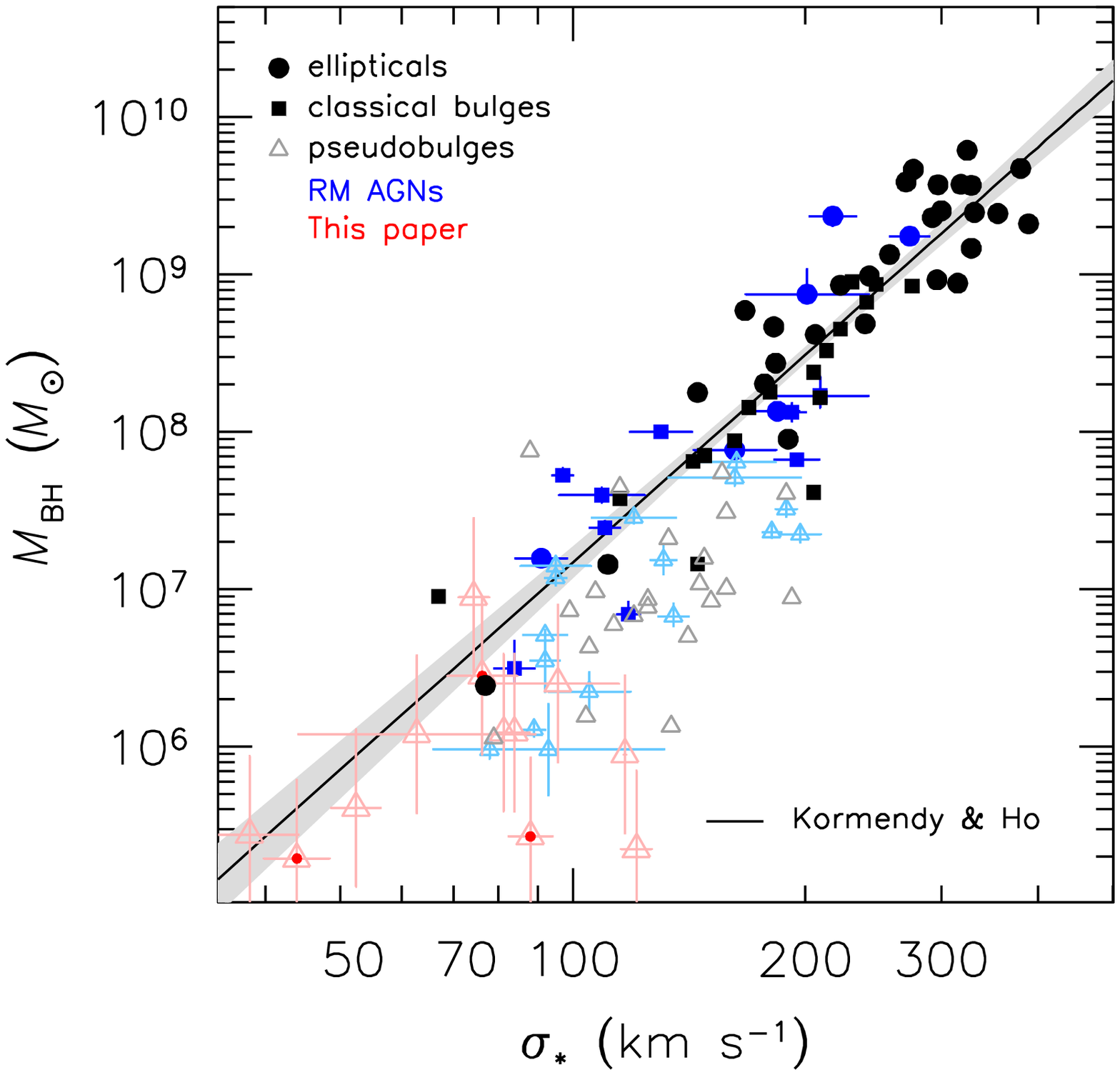,width=8.75cm,angle=0}
\figurenum{4}
\vskip 0.3cm
\figcaption[fig4.ps]{The $M_{\rm BH}-\sigma_*$ relation (solid line with
shaded 1$\sigma$ scatter) of inactive BHs detected via spatially resolved
dynamics (black symbols; Kormendy \& Ho 2013), of AGNs with BH masses measured
through reverberation mapping (blue symbols; Ho \& Kim 2014), and of the
low-mass active BHs (red symbols; this paper).  For the sake of clarity, error
bars are not shown for the inactive sample.  Classical bulges and ellipticals
are plotted in dark color; pseudobulges are plotted in a lighter shade.  We
assume that the low-mass objects are hosted by pseudobulges.  Their bulge
$\sigma_*$ are mostly estimated through the width of the core of 
\oiii\ \lamb 5007; the three objects with $\sigma_*$ directly measured using 
the \caii\ IR triplets are highlighted with a solid red point.
\label{fig4}}
\vskip 0.3cm

\noindent
their 
bandpass, and these features are detected clearly in all of them (Figure~2).  
We use the direct pixel-fitting method, as implemented in Ho et al. (2009), in 
conjunction with a K0~III velocity template star observed during the same 
observing run, to derive $\sigma_* = 43.9 \pm 4.3$ \kms\ for 
2XMMi~J032459.9$-$025612, $\sigma_* = 88.1 \pm 5.7$ \kms\ for 
2XMM~J134736.4+173404, and $\sigma_* = 76.3 \pm 7.9$ \kms\ for 
2XMM~J200824.5$-$444009.  The errors account for fitting uncertainties but not 
for uncertainties due to choice of velocity templates, as we did not observe a 
large number of stars.  However, comparison with the previous work of Xiao 
et al. (2011), who studied similar objects using MagE in exactly the same 
instrumental setup, verifies that our error estimates ($\sim 10\%$) are 
realistic.

\section{Broad \ha, Black Hole Masses, and Eddington Ratios}

We detect unambiguous broad \ha\ emission in all 12 sources and broad \hb\ 
emission in 8 out of 12, confirming that all are genuine type 1 AGNs.  
2XMM~J123103.2+110648 remains the only example of a pure type 2 source (Ho 
et al. 2012) in the sample of Kamizasa et al. (2012).  The broad \ha\ 
luminosities range from $\sim 10^{40}$ to $10^{42}$ \lum, with a median value 
of $2.5\times 10^{41}$ \lum, close to 1 order of magnitude less luminous than 
the peak of distribution of optically selected $z$ \lax\ 0.35 type~1 AGNs 
(Greene \& Ho 2007b), but similar to the subsample of type~1 AGNs with 
$M_{\rm BH} < 2 \times 10^6$ \solmass\ (Greene \& Ho 2007c).  Given the 
low-mass selection of our sample, it is not surprising that the widths of 
broad \ha\ are low (median FWHM = 934 \kms), well below the fiducial threshold 
of 2000 \kms\ for narrow-line Seyfert 1 galaxies (Osterbrock \& Pogge 1985).

The broad \ha\ measurements afford us the opportunity to reevaluate the BH 
masses and Eddington ratios of the sample.  

\begin{figure*}[t]
\centerline{\psfig{file=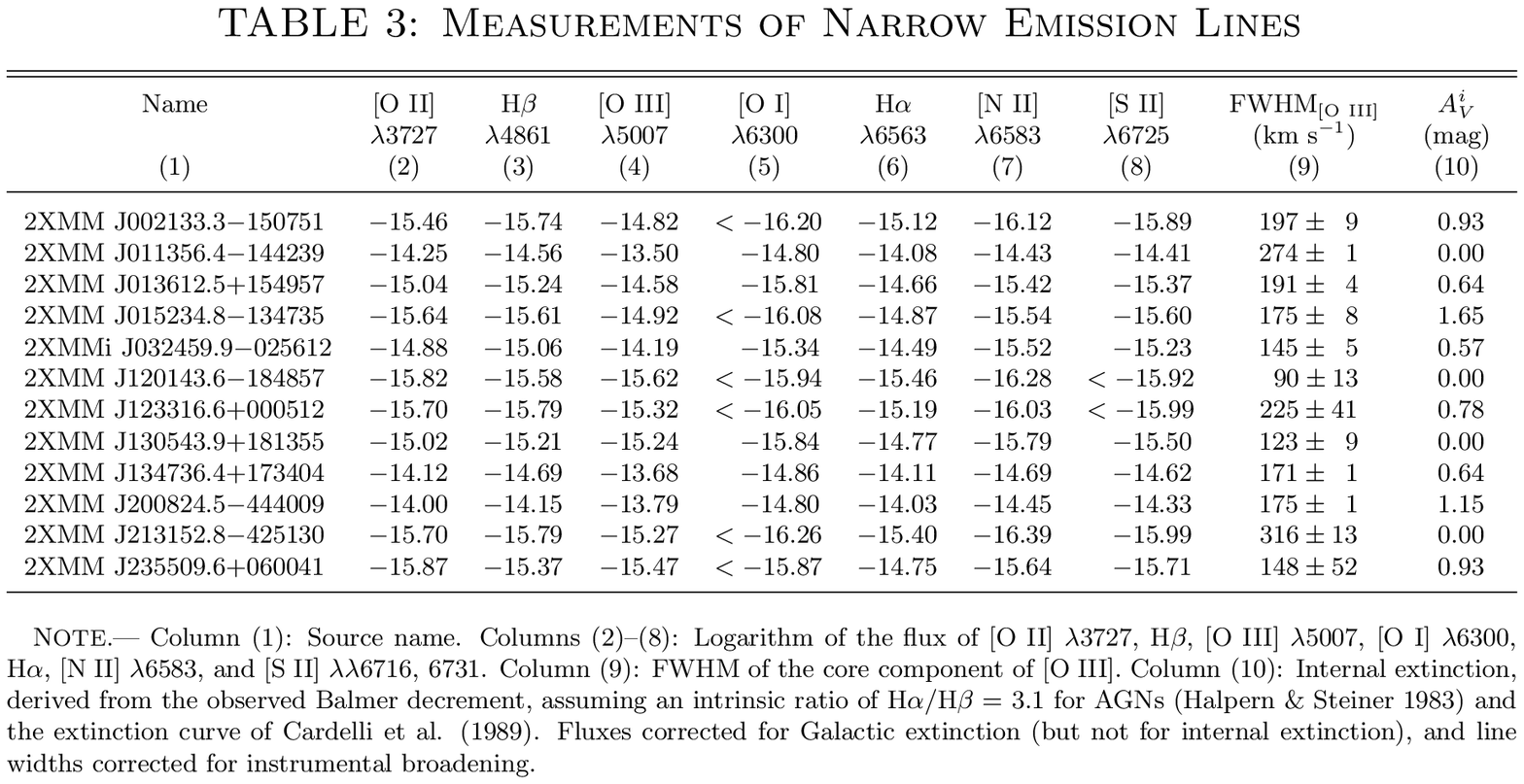,width=18.5cm,angle=0}}
\end{figure*}

\begin{figure*}[t]
\centerline{\psfig{file=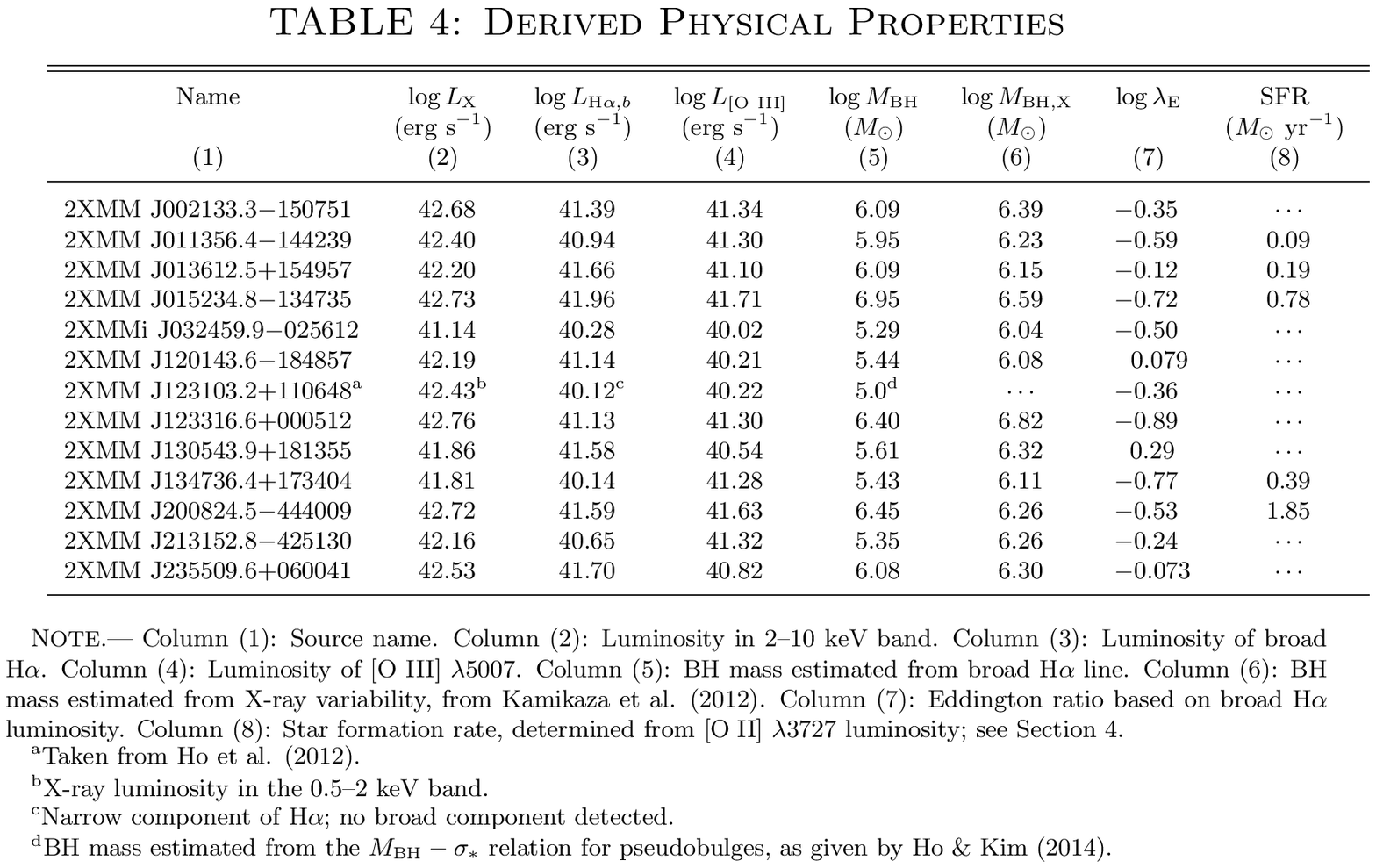,width=18.5cm,angle=0}}
\end{figure*}

\noindent
We estimate virial BH masses using
the formalism for single-epoch spectra, recently recalibrated for the \hb\ 
line by Ho \& Kim (2015), which takes into consideration the dependence of 
the virial coefficient (so-called $f$-factor) on bulge type (Ho \& Kim 2014), 
as well as the latest updates to the \msig\ relation of inactive galaxies 
(Kormendy \& Ho 2013), on which the calibration of the $f$-factor is based. 
As our sample specifically targets low-mass BHs, we expect, from experience 
with other  similar samples (Greene et al. 2008; Jiang et al. 2011b), that the 
host galaxies contain pseudobulges.  The BH mass estimator of Ho \& Kim (2015) 
is based on the FWHM of broad \hb\ and the AGN continuum luminosity at 5100 
\AA, whereas the primary measurements in this study come from the FWHM and 
luminosity of broad \ha.  Greene \& Ho (2005b) show that the 

\vskip 0.3cm
\figurenum{5}
\begin{figure*}[t]
\centerline{\psfig{file=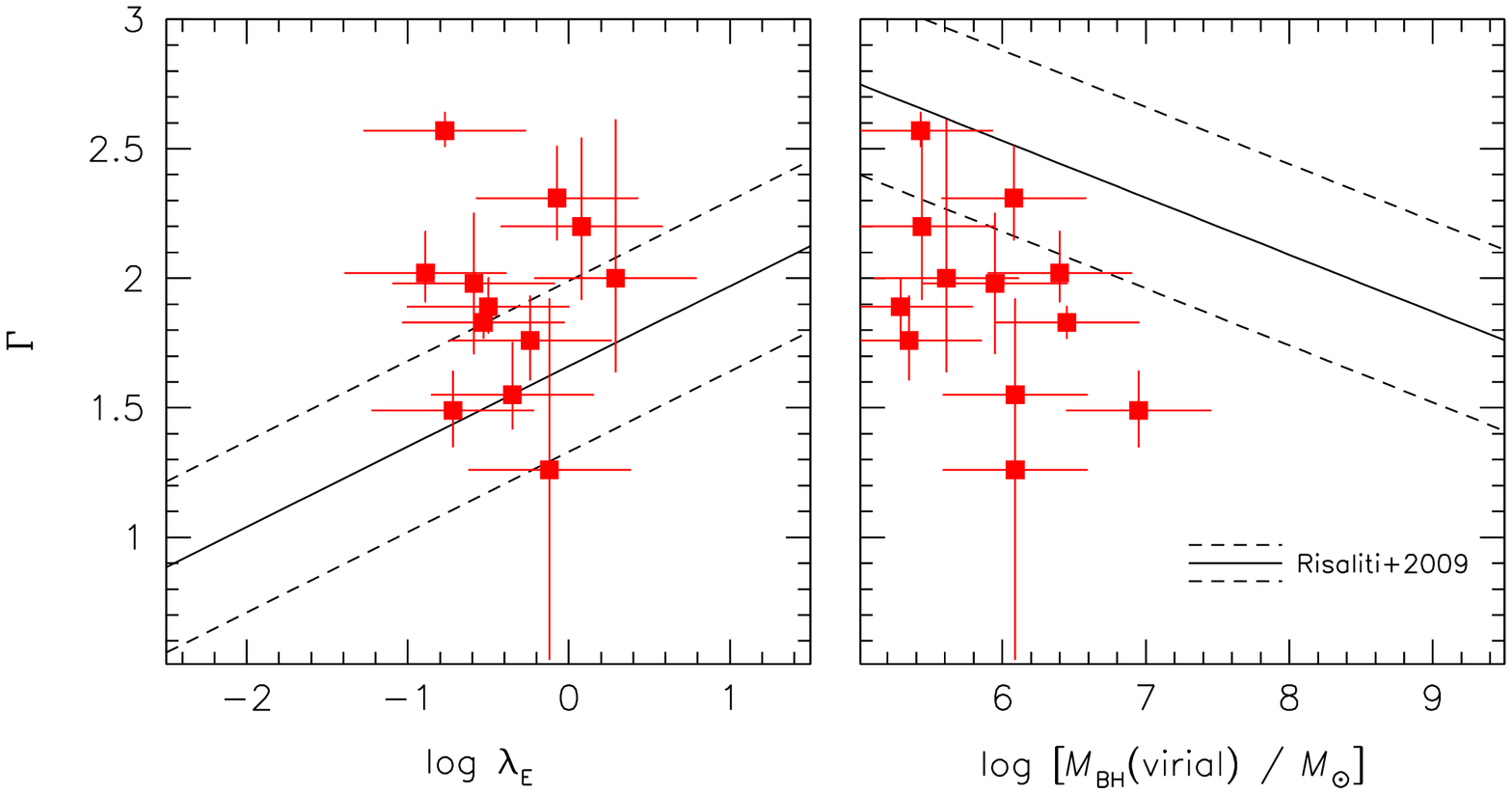,width=17.5cm,angle=270}}
\figcaption[fig5.ps]{Variation of X-ray photon index, $\Gamma$, calculated
over the 2--10 keV band (Kamizasa et al. 2012), with (left) Eddington ratio
and (right) BH mass.  The solid line is the best-fitting relation from
Risaliti et al. (2009), with the 1~$\sigma$ scatter denoted by the two dashed
lines.
\label{fig5}}
\end{figure*}
\vskip 0.3cm

\noindent
line widths of 
broad \ha\ and \hb\ scale closely with each other, as do the luminosities of 
the optical continuum and broad \ha.  Adopting the empirical correlations of 
Greene \& Ho (2005b, their equations 1 and 3), equation 4 of Ho \& Kim (2015) 
becomes

\begin{eqnarray}
\log \left(\frac{M_{\rm BH}}{M_\odot}\right) = 2.06\log \left(\frac{{\rm FWHM_{{\rm H}\alpha}}}{1000\,{\rm km\,s^{-1}}}\right) + \\
0.46\log \left(\frac{L_{{\rm H}\alpha,b}}{10^{42}\,{\rm erg\,s^{-1}}}\right) + 
k, \nonumber
\end{eqnarray}

\vskip 0.3cm
\noindent
where $k = 6.76$ for classical bulges and $k = 6.35$ for pseudobulges.  In this
paper, we adopt the zero point for pseudobulges.  The uncertainty on the 
masses is $\sim 0.5$ dex, based on the intrinsic scatter of the empirical
calibrations (Ho \& Kim 2014, 2015).

The new masses range from $M_{\rm BH} \approx 1.9\times 10^5$ to $8.3\times 
10^6$ \solmass, with a median value of $1.2\times 10^6$ \solmass.  This 
confirms that the selection method of Kamizasa et al. (2012), based on X-ray 
variability, indeed does isolate low-mass BHs effectively.  However, a direct 
comparison between their X-ray--based masses and ours (Figure~3) reveals 
large discrepancies.  Whereas the virial masses span nearly 2 dex, most of 
the X-ray--based masses are confined to a very narrow range of only $\sim 0.3$
dex.  Moreover, the scatter is not random: the X-ray--based masses are 
systematically higher than the virial masses, on average by $\sim 0.4$ dex.
While large uncertainties no doubt enter into the virial mass estimates, 
especially in the low-mass regime where the empirical calibrations remain 
poorly tested, we have reasonable confidence that our derived masses are not 
affected by large, systematic biases because they more-or-less follow the 
\msig\ relation, at least within the relatively large scatter in the 
pseudobulge regime of the correlation (Figure~4).  This has been established 
for other samples of low-mass type~1 AGNs (Barth et al. 2005; Greene \& Ho 
2006b; Xiao et al. 2011; see also Jiang et al. 2011a).  For the present 
analysis, we only have direct stellar velocity dispersions for three sources 
(Section~2); as for the rest, we assume $\sigma_*={\rm FWHM_{[O~III]}/2.35}$ 
(appropriate for a Gaussian profile), where ${\rm FWHM_{[O~III]}}$ pertains 
to the core of the line.  This approximation has been shown to be valid 
(Nelson \& Whittle 1996; Greene \& Ho 2005a), especially for lower-ionization 
lines such as \sii\ or \nii\ (Ho 2009).  The \sii\ and \oiii\ lines of our 
sample have very similar FWHM values, and so we choose the measurements based 
on the stronger \oiii\ line.  

The offset between the virial and X-ray--based BH masses can be compensated by adopting in equation 1 the zero point for classical bulges instead of pseudobulges, which differs by precisely this amount ($\sim 0.4$ dex).  However, as discussed above, we have strong reason to believe that these low-mass AGNs are hosted by pseudobulges. 

Next, we convert the broad \ha\ luminosities to bolometric luminosities, using 
the $L_{{\rm H}\alpha,b}-L_{\rm 5100~\AA}$ relation of Greene \& Ho (2007c) 
and a canonical bolometric correction of $L_{\rm bol} = 9.8L_{\rm 5100~\AA}$ 
(McLure \& Dunlop 2004).  The resulting Eddington ratios are surprisingly 
high ($\lambda_{\rm E} \approx 0.1-2$; median $\lambda_{\rm E} = 0.44$), 
not dissimilar from the optically selected sample of Greene \& Ho (2007c). 
The bolometric luminosity can be estimated alternatively from the available 
X-ray luminosity, using a luminosity-dependent X-ray bolometric correction 
factor (Ho 2008).  Adopting  $L_{\rm bol} = 20 L_{\rm X}$ for the 2--10 keV 
band (Vasudevan \& Fabian 2007), we find consistent values of $\lambda_{\rm E}
\approx 0.1-1$, with a median value of $\lambda_{\rm E} = 0.38$.  

We note, in agreement with Kamizasa et al. (2012), that the hard X-ray (2--10 
keV) photon index does not correlate with BH mass or Eddington ratio in the 
manner predicted from extrapolation of AGN samples of higher mass and 
luminosity (Fig.~5).

\vskip 0.3cm
\psfig{file=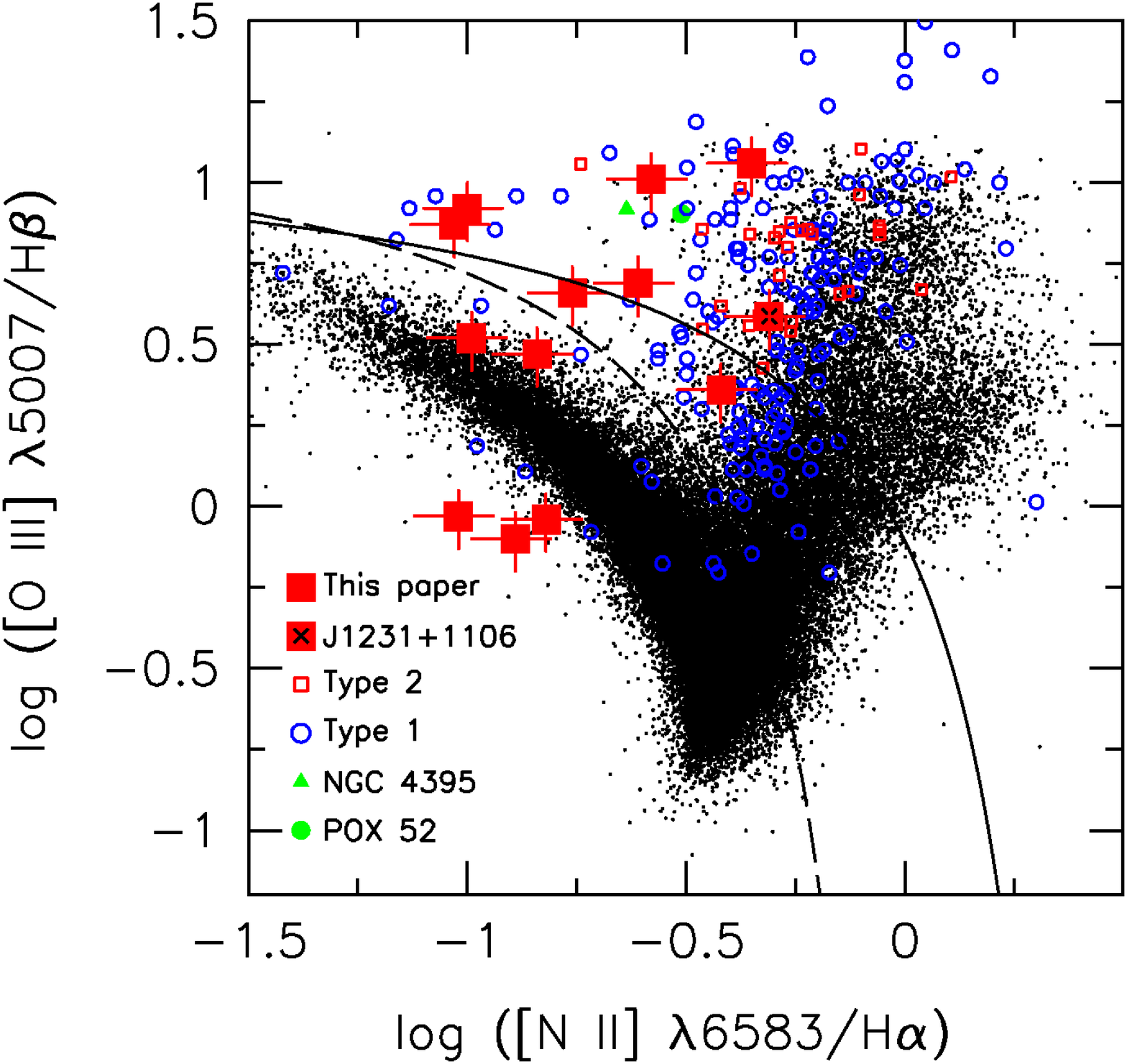,width=8.75cm,angle=0}
\figurenum{6}
\vskip 0.3cm
\figcaption[fig6.ps]{
Line-ratio diagnostic diagram of \oiii\ \lamb5007/\hb\ versus  \nii\
\lamb6583/\ha.  The small black points are Sloan Digital Sky Survey 
measurements from Kauffmann et al. (2003) with S/N larger
than 6.  The new sample of 12 objects studied in this paper are plotted as 
large red squares; J1231+1106, from Ho et al. (2012), is marked with an 
additional cross.  Conservative errors of $\sim 20\%$ are assigned to our 
line ratio measurements.  Small open red squares are low-mass type~2 Seyferts 
from Barth et al. (2008), while the type~1 sample of Greene \& Ho (2007c) is 
shown as open blue circles.  NGC~4395 and POX~52 (Barth et al. 2004) are 
plotted as a filled green triangle and circle, respectively.  Star-forming 
galaxies are located to the left of the dashed line, AGNs are found to the 
right of the solid line, and composite sources lie in between (Kewley et al. 
2006).
\label{fig6}}
\vskip 0.3cm

\section{Constraints on Star Formation Rate}

The spectra exhibit a rich set of narrow emission lines generally consistent
with AGN photoionization.  On the
\oiii\ \lamb5007/\hb\ versus  \nii\ \lamb6583/\ha\ diagnostic diagram 
(Figure~6), most of the objects lie in the region intermediate between 
high-excitation star-forming galaxies and classical Seyferts, a manifestation 
of the relatively low metallicity associated with the lower-mass host galaxies 
in our study (Ludwig et al. 2012).  Our sources overlap well with other 
samples of type~1 (Barth et al. 2004; Greene \& Ho 2004, 2007c) and type~2 
(Barth et al. 2008) low-mass AGNs.  As was found for the Greene \& Ho (2007c) 
sample, some of the sources in our sample (5/13) have narrow-line intensity 
ratios that formally place them in the territory of star-forming galaxies.   
We suspect that star formation contaminates the integrated spectra of these 
objects, and possibly others in our sample.  

Can the star formation rate (SFR) be constrained in these systems?  Ho (2005) 
proposed that the strength of \oii\ \lamb3727, normally a widely used SFR 
indicator in star-forming galaxies at intermediate redshifts, can also be 
utilized for this purpose in AGNs.  The fundamental premise behind this idea 
is that, while the narrow-line regions of AGNs certainly also emit \oii, the 
level of \oii\ emission in high-excitation sources (i.e. Seyferts and quasars)
is well-constrained by the level of \oiii\ \lamb5007, which can be directly 
observed.  The intensity ratio \oii/\oiii, which varies strongly with the 
ionization parameter, decreases with increasing \oiii\ luminosity (Kim et al. 
2006).  To zeroth order, all of the \oiii\ emission in luminous AGNs comes 
from nonstellar photoionization.  Star 

\psfig{file=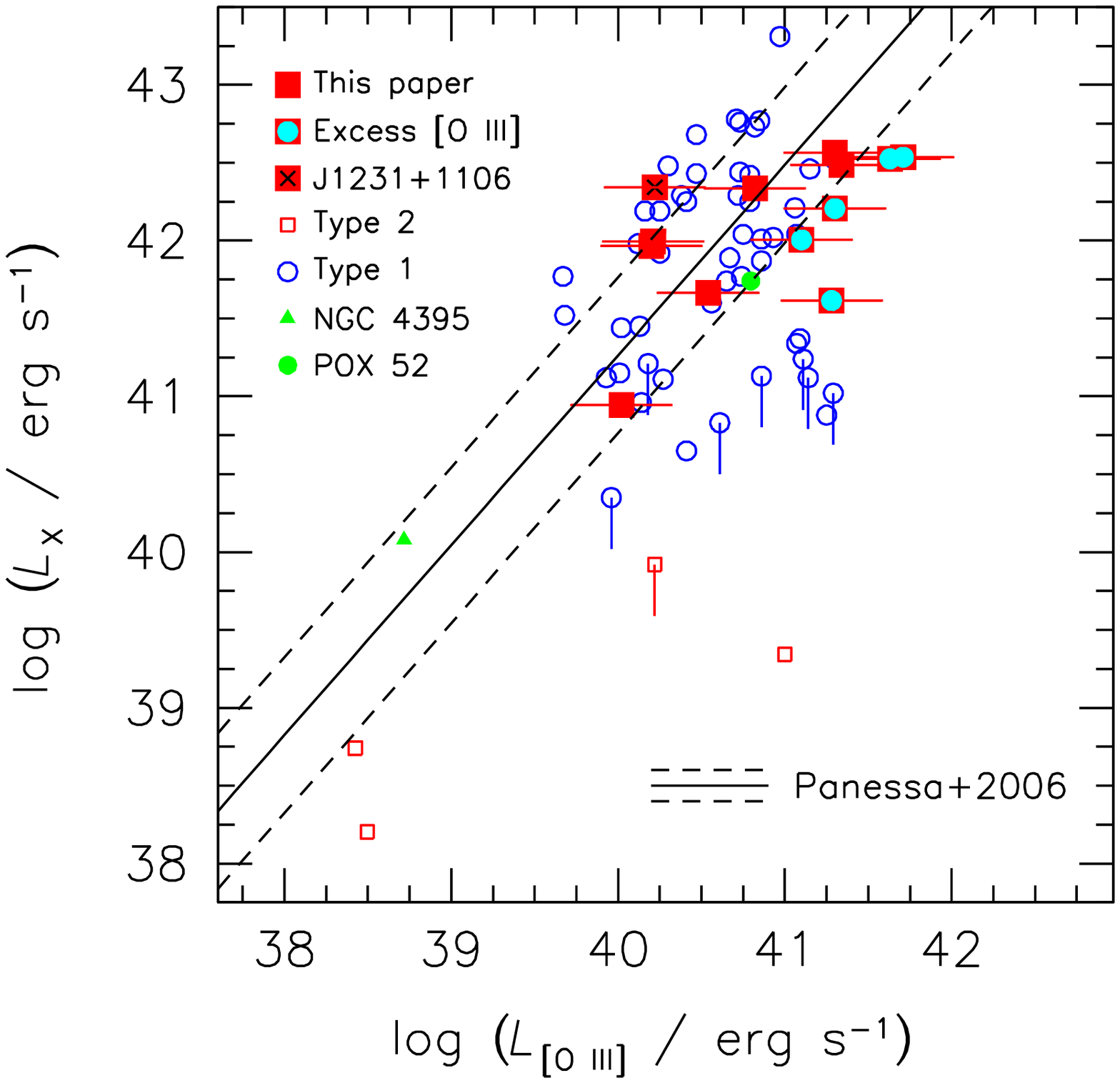,width=8.75cm,angle=0}
\figurenum{7}
\vskip 0.3cm
\figcaption[fig7.ps]{
Correlation between intrinsic soft X-ray luminosity in the 0.5--2 keV band and
extinction-corrected \oiii\ \lamb5007 luminosity for the 12 objects in this
paper (filled red squares), 2XMM~J123103.2+110648 (filled red square with
cross; Ho et al. 2012), low-mass type 2 AGNs (open red squares; Thornton et
al. 2009); low-mass type 1 AGNs (open blue circles; Greene \& Ho 2007a;
Desroches et al. 2009; Dong et al. 2012a), NGC~4395 (filled green triangle;
Panessa et al. 2006, adjusted to a distance of 4.3 Mpc), and POX~52 (solid
green circle; Thornton et al. 2008).  The solid line is the best-fitting
relation from Panessa et al. (2006), with the 1~$\sigma$ scatter denoted by
the two dashed lines; their X-ray luminosities were transformed from the 2--10
keV band to the 0.5--2 keV band assuming a photon index of $\Gamma = 1.8$.  For the objects
studied in this paper and in Ho et al. (2012), we conservatively assign a
uncertainty of 0.3 dex for the $L_{\rm [O~III]}$.  The five objects with 
``excess" \oiii\ emission are highlighted with a cyan center.
\label{fig7}}
\vskip 0.3cm

\noindent
formation, if present, contributes 
negligibly to \oiii\ because the host galaxies, being massive in general for 
luminous Seyferts and quasars, should have high metallicity (Tremonti et al. 
2004).  Thus, given an observed strength of \oiii\ in an AGN, its associated 
level of \oii\ emission can be predicted, and any ``excess'' beyond that can 
be attributed to star formation (Ho 2005; Kim et al. 2006).

The \oii-based technique to measure SFRs in AGNs depends on a critical 
condition: that the host galaxy has high metallicity.  Unfortunately, this 
requirement breaks down for low-mass galaxies, and, at sufficiently high 
redshifts, even for high-mass galaxies.  O-star photoionization of 
low-metallicity gas produces high-excitation \hii\ regions, which, like AGNs, 
emit strong \oiii\ \lamb 5007 (Osterbrock 1989).  Here, we suggest that this 
difficulty can be circumvented with the help of an additional constraint from 
X-ray observations.  The extinction-corrected \oiii\ luminosity of Compton-thin
AGNs correlates well with the intrinsic X-ray luminosity (Bassani et al. 1999),
over a wide range of luminosities and masses (Panessa et al. 2006; Stern \& 
Laor 2012), including optically selected low-mass AGNs (Dong et al. 2009; Ho 
et al. 2012).  Hence, provided that a robust X-ray detection is available, we 
can deduce roughly the corresponding \oiii\ luminosity that {\it should}\ 
be associated with the AGN, and thereby also its fractional contribution to 
the \oii\ emission.

Figure~7 confirms that some of the objects in the current sample of AGNs 
selected by X-ray variability indeed do depart systematically from the 
canonical $L_{\rm X}-L_{\rm [O~III]}$ correlation.  For 

\psfig{file=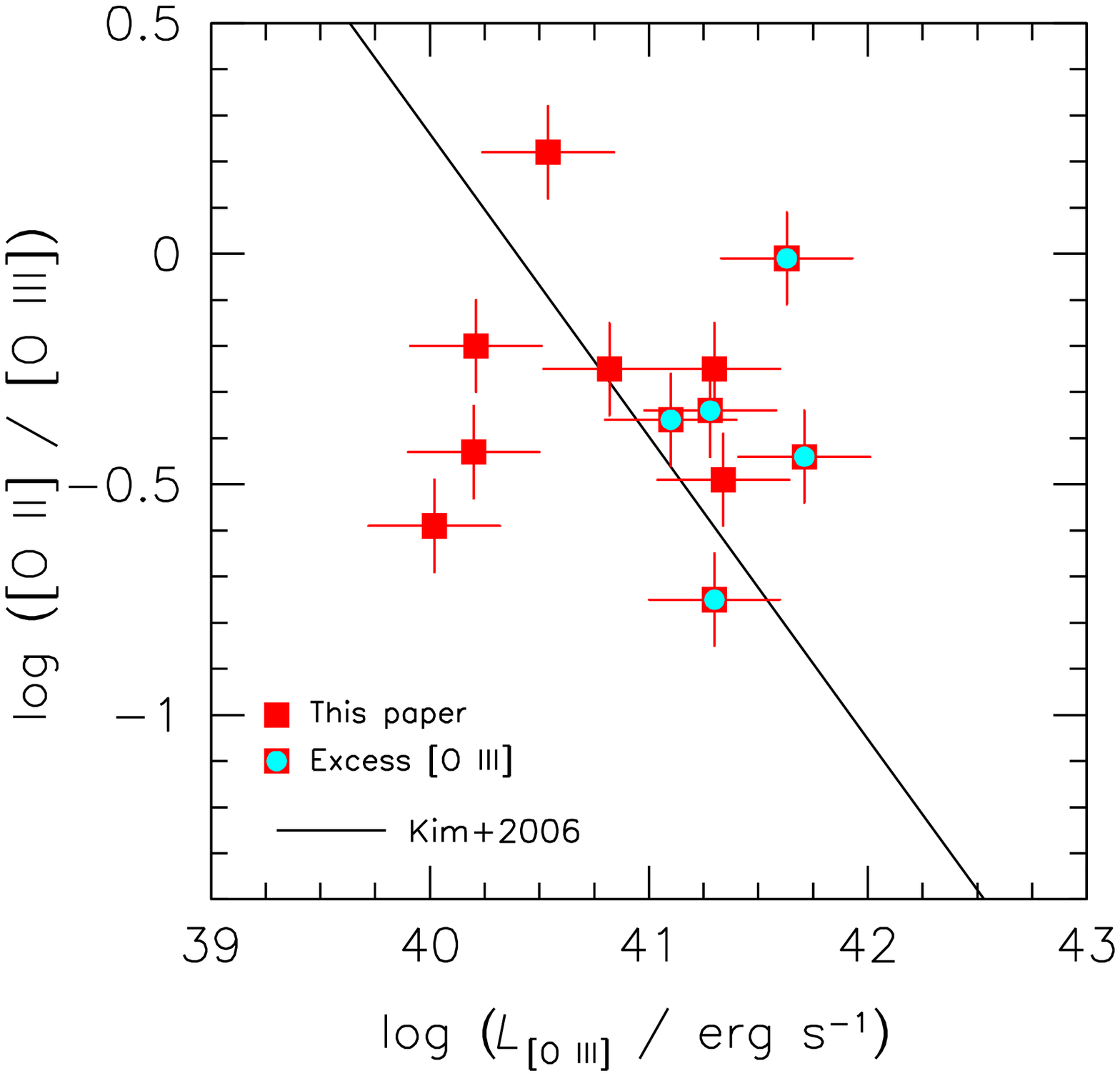,width=8.75cm,angle=0}
\figurenum{8}
\vskip 0.3cm
\figcaption[fig8.ps]{
Dependence of \oii/\oiii\ ratio on extincton-corrected \oiii\ luminosity. The 
solid line is the empirical relation derived from a large sample of type~1 
AGNs selected from the Sloan Digital Sky Survey (Kim et al. 2006).  Most of 
the sample lies systematically above the relation, expecially the five sources 
that deviate the most from the $L_{\rm X}-L_{\rm [O~III]}$ correlation 
(Figure~7), in the sense that at a given \oiii\ luminosity there is ``extra'' 
\oii\ emission, which we argue originates from and hence traces star 
formation in the host galaxy.
\label{fig8}}
\vskip 0.3cm

\noindent
illustrative purposes,
we highlight (red squares with cyan centers) the five objects that formally 
lie outside the 1$\sigma$ scatter of the $L_{\rm X}-L_{\rm [O~III]}$ relation 
of Panessa et al. (2006), in the sense that at a given X-ray luminosity they 
are overluminous in \oiii\ by a factor of $\sim 3$.  This is to be expected, 
if, as we suspect, high-excitation \hii\ regions contribute to the integrated 
\oiii\ emission in our sample of low-mass AGNs, which most likely live in 
low-mass, low-metallicity host galaxies.  These very same five objects stand 
out as having systematically higher \oii/\oiii\ ratios for their \oiii\ 
luminosities (Figure~8), indicating that star formation contributes to their 
\oii\ emission. (Interestingly, however, they do not stand out in any obvious 
way in the traditional dignostic diagram in Figure 6.)

We apply the above procedure to derive ``AGN-decontaminated'' \oii\ 
luminosities for our objects.  As in Ho (2005), we use the empirical 
calibration of Kewley et al. (2004) to derive SFRs from $L_{\rm [O~II]}$, 
adopting, for concreteness, a metallicity of 0.5 solar.  Most objects have 
rather modest SFRs (\lax\ 1 \solmass\ \peryr; Table~4).  

\section{The Peculiar Narrow-line Kinematics of 2XMM~J200824.5$-$444009}

We draw attention to the highly unusual kinematics of the narrow emission
lines in 2XMM~J200824.5$-$444009 (Figure~1).  The profile of \oiii\ \lamb\lamb 
4959, 5007 qualifies the source as ``double-peaked'', but in other objects of 
this class the two components typically have comparable widths (e.g., Liu 
et al. 2010; Smith et al. 2010; Shen et al. 2011; Comerford et al. 2012; 
Ge et al. 2012).  Here, the two components have very different profiles: a 
sharp, core with FWHM = 175 \kms\ plus an extremely broad component with FWHM 
= 870 \kms\ blueshifted by 381 \kms.  This characteristic double-component 
structure is clearly present in \hb, and, although less obvious visually, 
also in \sii.  It is crucial to properly account for this complex narrow-line 
profle to deblend narrow \ha\ and \nii, in order to obtain an accurate 
measurement of broad \ha.

\section{Summary}

AGNs with \mbh\ \lax\ $10^6$ \solmass\ are an important constituent for 
understanding the full demography of central BHs.  The vast majority of 
currently known low-mass AGNs derive from optical selection.  In a recent 
study, Kamizasa et al. (2012) presented a sample of candidate low-mass BHs 
based on X-ray variability characteristics.  We used the Magellan 6.5m Clay 
telescope to obtain high-quality, optical echellette spectra of 12 out of the 
15 sources from Kamizasa's 
sample.  Broad \ha\ emission is unambiguously detected in all the objects, 
permitting us to calculate robust virial BH masses and Eddington ratios.  We 
confirm that the sample contains low-mass BHs accreting at a high fraction of 
their Eddington rate: median \mbh\ $\approx$ $1.2\times10^6$ \solmass; median 
$\lambda_{\rm E}=0.44$.  Assuming the AGNs to be hosted by pseudobulges, our BH masses are significantly and 
systematically larger than those estimated from X-ray variability.  The 
moderately high resolution of our spectra allow us to measure the central 
velocity dispersion of the host galaxies, either directly from stars or 
indirectly from ionized gas.  We show that the virial BH masses follow the 
\msig\ relation of pseudobulges.  Lastly, we argue that at least some of the
objects have ongoing star formation, and we introduce a method to estimate 
the SFRs of low-metallicity AGN host galaxies based on measurements of X-rays
and optical emission lines.

\acknowledgements
LCH acknowledges financial support from Peking University, the Kavli Foundation, the Chinese Academy of Science through grant No. XDB09030102 (Emergence of Cosmological Structures) from the Strategic Priority Research Program, and from the National Natural Science Foundation of China through grant No. 11473002.  We are grateful to George Becker for providing the data reduction pipeline for MagE.


\end{document}